\documentclass[9pt,twocolumn,twoside]{osajnl}

\journal{ol} 

\setboolean{shortarticle}{true}

\usepackage{lineno}
\usepackage{textcomp}
\usepackage{multirow}
\usepackage{xcolor}

\title{Integrated lithium niobate intensity modulator on a silicon handle with slow-wave electrodes}

\author[1,2*]{Sean Nelan}
\author[2]{Andrew Mercante}
\author[1,2]{Shouyuan Shi}
\author[2]{Peng Yao}
\author[2]{Eliezer Shahid}
\author[2]{Benjamin Shopp}
\author[1,2]{Dennis W. Prather}

\affil[1]{School of Electrical and Computer Engineering, University of Delaware, Newark, Delaware 19716, USA}
\affil[2]{Phase Sensitive Innovations, Newark, Delaware 19713, USA}

\affil[*]{Corresponding author: snelan@udel.edu}

\begin{abstract}

Segmented, or slow-wave electrodes have emerged as an index-matching solution to improve bandwidth of traveling-wave Mach Zehnder and phase modulators on the thin-film lithium niobate on insulator platform. However, these devices require the use of a quartz handle or substrate removal, adding cost and additional processing. In this work, a high-speed dual-output electro-optic intensity modulator in the thin-film silicon nitride and lithium niobate material system that uses segmented electrodes for RF and optical index matching is presented. The device uses a silicon handle and does not require substrate removal. A silicon handle allows the use of larger wafer sizes to increase yield, and lends itself to processing in established silicon foundries that may not have the capability to process a quartz or fused silica wafer. The modulator has an interaction region of 10 mm, shows a DC half wave voltage of 3.75 V, an ultra-high extinction ratio of roughly 45 dB consistent with previous work, and a fiber-to-fiber insertion loss of 7.47 dB with a 95 GHz 3 dB bandwidth.
\end{abstract}

\setboolean{displaycopyright}{true}

\begin{document}

\maketitle

\section{Introduction}

Global networking traffic continues to grow past the capabilities of current infrastructure while existing systems become slow, inefficient and unreliable \cite{alvarado_impact_2016,zhang_integrated_2021,yue_experimental_2019,nelan_compact_2022}. As coaxial transmission lines are upgraded to support the next communication standard and cellular towers reinforced to support their weight, a band-aid is applied to a deeper issue: existing hardware that supports these systems has reached the limit of its capabilities and must be replaced not only to maintain stable operation, but to raise the technological "ceiling" and pave the way for the faster, more reliable and more secure networks of the future. Fiber-optic cables can replace the heavy, lossy, and band-limited transmission lines of today's networks, but cannot be simply "plugged-in" to an existing antenna, networking hub, or cellular tower. We must introduce a "converter" that can read an RF signal meant for a coaxial cable and convert it to an optical signal that can be transmitted through a fiber-optic cable. The optical modulator provides this capability \cite{wooten_review_2000}. Two approaches to optical modulation have risen to prominence in the last two decades: free-carrier dispersion-based modulation and electro-optic (EO) modulation\cite{zhang_integrated_2021,wooten_review_2000}. Free-carrier dispersion-based modulation involves the manipulation of free-carriers in a semiconductor to change the real and imaginary refractive index of the material, while electro-optic modulation takes advantage of the non-linear optical properties of a crystal to modulate a beam of light within an applied electric field \cite{ghione_2009,zhang_integrated_2021}. EO modulators show increased bandwidth, lower voltage operation, increased extinction ratio (ER), higher spurious-free dynamic range, zero chirping and increased power handing capability over their free-carrier dispersion counterparts \cite{ghione_2009,zhang_integrated_2021,nelan_compact_2022,ahmed_high-efficiency_2020,ahmed_subvolt_2020,wooten_review_2000}. Thin-film lithium niobate (LiNbO$_\mathrm3$) on insulator (TFLNOI) is chosen as the electro-optic medium for it's high second-order non-linearity ($\chi^{(2)}$), low third-order non-linearity ($\chi^{(3)}$), low optical absorption, and exceptional environmental stability \cite{ghione_2009,zhang_integrated_2021,nelan_compact_2022}. TFLNOI offers tighter mode confinement leading to a smaller footprint, and improved voltage-bandwidth performance over bulk LiNbO$_\mathrm3$ devices \cite{ahmed_subvolt_2020,zhang_integrated_2021,kharel_breaking_2021}. To guide a tightly-confined optical mode and support the use of an on-chip fiber-coupler at the end facets of the device, silicon nitride (SiN$_\mathrm{x}$) is chosen as the ideal candidate to form a low-loss strip-loaded waveguide on the TFLNOI \cite{nelan_compact_2022,ahmed_high-efficiency_2020}.

\begin{figure*}[ht]
\centering
{\includegraphics[width=\linewidth]{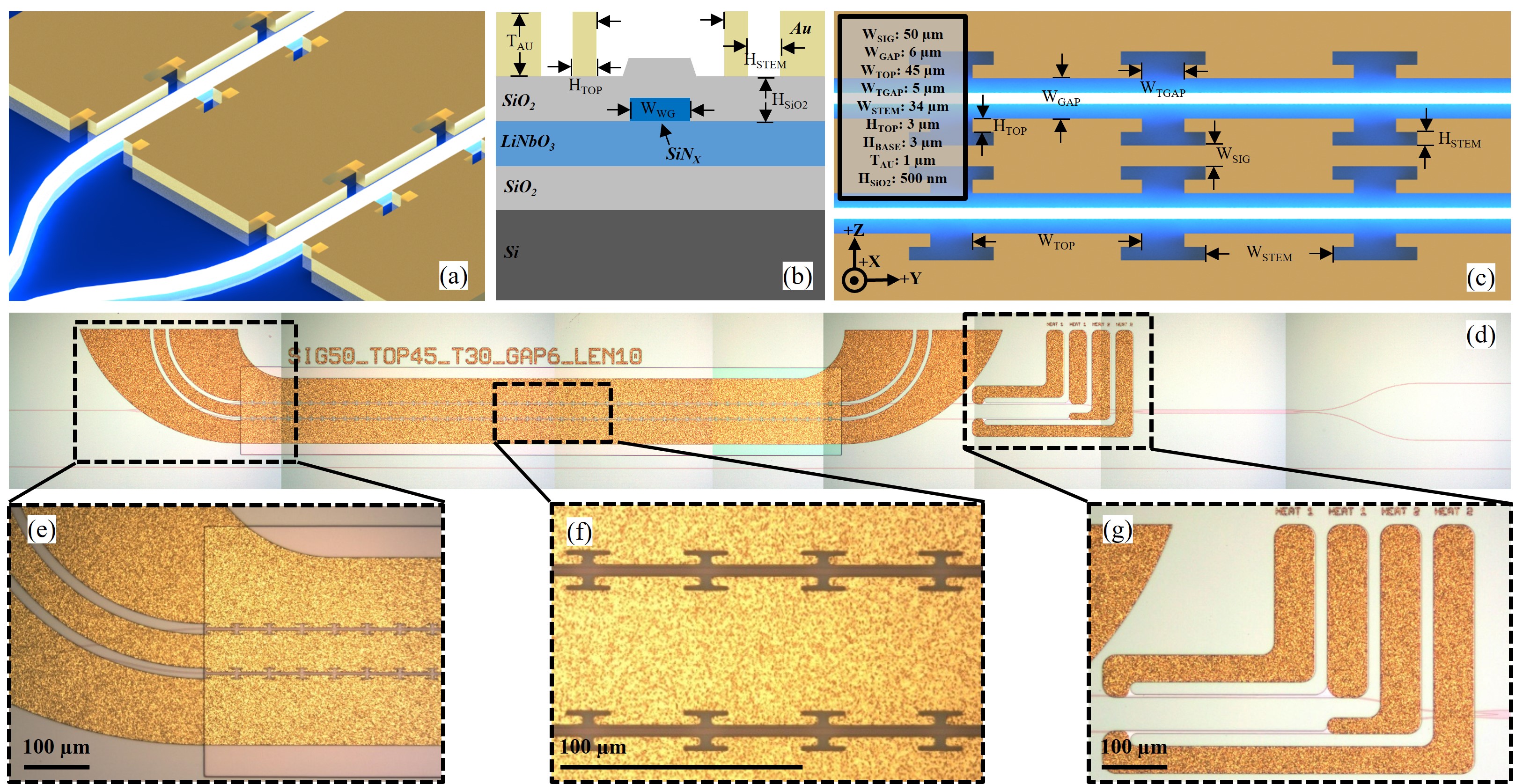}}
\caption{(a) Artistic overview of the segmented electrode structures. (b) Cross-sectional view of the material stack and dimension definitions. (c) Top-down view of the segmented-electrodes with dimension definitions. (d) Top-down view of the fabricated device. Optical and RF propagation occurs from left to right. (e) Detail view of the transition from the probe-region to interaction region. (f) Detail view of the interaction region showing to-scale segmented electrodes. (g) Detail view of the thermal bias heaters.}
\label{fig:overview}
\end{figure*}

In previous work, co-planar waveguide (CPW) electrodes have been used to apply the RF modulating signal \cite{nelan_compact_2022,ahmed_subvolt_2020,ahmed_high-efficiency_2020}. To closely match the RF phase and optical group index ($n_{RF}$ and $n_{og}$), an index matching epoxy (Masterbond™ UV15) is applied to the interaction region between the signal and ground electrodes \cite{ahmed_high-efficiency_2020,nelan_compact_2022}. However, the presence of the epoxy nearly doubles the RF loss ($\alpha_{RF}$) over 100 GHz compared to a standard, unmatched CPW \cite{nelan_compact_2022}. Moreover, it is not possible to precisely tune the RF index when using an index matching fluid, where a mismatch of even 1-2$\%$ between $n_{og}$ and $n_{RF}$ can greatly reduce the 3 dB bandwidth of the device \cite{zhang_integrated_2021,ghione_2009,kharel_breaking_2021}. Finally, UV15 is not environmentally stable, and may breakdown at high temperatures or in the presence of ultra-violet (UV) radiation. Micro-structured, or "segmented" electrodes have been used in literature to slow the RF phase velocity and match the RF and optical index \cite{kharel_breaking_2021,chen_high_2022,li_analysis_2004,motta_design_2017,jaehyuk_shin_conductor_2011,huang_advanced_2021}. This provides tunability of $n_{RF}$, and drastically reduces the RF loss compared to a UV15-clad electrode, increasing EO bandwidth. However, these require either the use of a quartz handle or a complicated substrate-removal technique after electrode fabrication to match $n_{og}$ and $n_{RF}$\cite{kharel_breaking_2021,chen_high_2022}. These methods then preclude fabrication of the device in any established silicon foundry. In this work, a dual-output Mach Zehnder modulator (MZM) using micro-structured CPW electrodes is fabricated on a TFLNOI wafer with a silicon handle without the use of substrate removal or back-dicing techniques. An artist's rendition of the segmented electrodes is seen in Fig. \ref{fig:overview}. (a). A silicon handle allows the use of larger wafer sizes to increase yield and lends itself to processing in established silicon foundries that may not have the capability to process a quartz or fused silica wafer, or to etch through a LiNbO$_\mathrm3$ layer. The modulator has an interaction region of 10 mm, shows a DC half wave voltage (V$\pi$) of 3.75 V, an ultra-high ER of roughly 45 dB, and a fiber-to-fiber insertion loss of 7.47 dB. The measured 3 dB bandwidth is roughly 95 GHz. To the best of our knowledge, this is the first time a traveling-wave EO modulator has been designed, fabricated and characterized with slow-wave segmented electrodes in the LiNbO$_\mathrm3$ on Si material platform without substrate removal. 

\section{Design and Fabrication}

An optical modulator should maintain a small footprint and a wide RF bandwidth if it is to be used in the telecommunications system of the future \cite{wooten_review_2000,zhang_integrated_2021}. To achieve this, the traveling-wave EO modulator must have minimal RF loss and near-perfect index matching between $n_{og}$ and $n_{RF}$ \cite{kharel_breaking_2021}. The introduction of "T"-shaped segments along the interaction region of the electrode enables tuning of the RF phase velocity for better matching of $n_{og}$ and $n_{RF}$ through periodic capacitive loading, and can decrease current crowding within the electrode gap \cite{motta_design_2017,li_analysis_2004,kharel_breaking_2021,jaehyuk_shin_conductor_2011}. Recently, segmented electrodes have been designed and fabricated on quartz (Qz) and substrate-removed Si handles, and before that on III-V compound semi-conductor modulators to slow the RF traveling wave \cite{kharel_breaking_2021,chen_high_2022,li_analysis_2004}. Here, we employ segmented electrodes on a Si handle without substrate removal to raise $n_{RF}$ to ~2.134 to match $n_{og}$ = 2.138, where $n_{RF}$ of a standard CPW in this material system is ~1.92. Considering the relation between $n_{RF}$ and the microwave impedance of the loaded line (Z$_{l}$) given by Eq. \ref{eq:inductcap}, we are able to tune the slowing of the wave by adjusting W$_\mathrm{STEM}$ to increase or decrease the inductance of the CPW by changing the effective distance that the current must travel along the electrode \cite{li_analysis_2004,motta_design_2017}.
\begin{equation}
Z_{l} = \sqrt{\frac{ L_{l}}{C_{l}}},\quad n_{RF}=c_{0}\sqrt{L_{l}C_{l}}
\label{eq:inductcap}
\end{equation}
Here, Z$_l$, L$_l$ and C$_l$ are the microwave impedance, inductance and capacitance of the loaded CPW, respectively, and c$_0$ is the speed of light. Considering the 500 nm height of the SiO$_2$ top cladding (T$_\mathrm{SIO2}$) under the Au electrode, the electrode gap width (W$_\mathrm{GAP}$) is set to 6 $\mu$m to avoid unwanted optical absorption loss from the optical mode interacting with the Au electrode. The voltage-length product (V$\pi$-L) with different H$_\mathrm{CLAD}$ and W$_\mathrm{GAP}$ values is shown in Fig. \ref{fig:vpidata}. (a). These values can be adjusted to meet specific design goals \cite{ghione_2009,ibarra_fuste_bandwidthlength_2013}. A lower V$\pi$-L can be achieved at the cost of RF and optical propagation loss when W$_\mathrm{GAP}$ and H$_\mathrm{CLAD}$ are made smaller, or by using a ridge-etched LiNbO$_\mathrm3$ waveguide at the cost of excess coupling loss. The height of the top of the segment (H$_\mathrm{TOP}$) is set to 3 $\mu$m, where RF loss may increase with a larger dimension. With W$_\mathrm{TOP}$ = 45 $\mu$m and W$_\mathrm{TGAP}$ = 5 $\mu$m to reduce the dispersion effect and raise the Bragg cutoff frequency above 1 THz, W$_\mathrm{STEM}$ is experimentally chosen to be 34 $\mu$m to achieve $n_{RF@120GHz}$ = 2.134. Moreover, by varying the width of W$_\mathrm{STEM}$ from 10 $\mu$m to 42 $\mu$m, and consequentially the inductance of the electrode, we were able to adjust the RF index from $n_{RF}$ = 2.3 to $n_{RF}$ =  2.1, proving the effectiveness of the design on different material systems. A comparison between measured L$_l$ and C$_l$ of the segmented and standard CPW is shown in Fig. \ref{fig:hfdata}. (f). Finally, H$_\mathrm{STEM}$ and W$_\mathrm{SIG}$, and T$_\mathrm{AU}$ are set to 3 $\mu$m, 50 $\mu$m and 1 $\mu$m, respectively, to reduce RF loss and most closely match Z$_l$ to the source impedance Z$_0$ = 50 $\Omega$. The effect of each dimension are shown in Table \ref{tab:segdimensions}.

\begin{table}[h]
\centering
\caption{\bf Electrode Dimension vs. RF Index and Loss}
\begin{tabular}{c c c c}
\hline
Dimension&Size&RF Index ($n_{RF}$)&RF Loss ($\alpha_{RF}$)\\
 \hline
 W$_\mathrm{SIG}$ & $\uparrow$ & $\uparrow$ & $\downarrow$\\
 W$_\mathrm{GAP}$ & $\uparrow$ & $\downarrow$ & $\downarrow$\\
 W$_\mathrm{TOP}$ & $\uparrow$ & $\uparrow$ & --\\
 W$_\mathrm{TGAP}$ & $\uparrow$ & -- & --\\
 W$_\mathrm{STEM}$ & $\uparrow$ & $\downarrow$ & $\uparrow$\\
 H$_\mathrm{TOP}$ & $\uparrow$ & $\uparrow$ & $\uparrow$\\
 H$_\mathrm{BASE}$ & $\uparrow$ & $\uparrow$ & $\uparrow$\\
 T$_\mathrm{AU}$ & $\uparrow$ & $\downarrow$ & $\downarrow$\\
 \hline
\end{tabular}
  \label{tab:segdimensions}
\end{table}

The Si handle has a relatively high permittivity ($\epsilon_\mathrm{Si}$ = 11.7) compared to Qz ($\epsilon_\mathrm{Qz}$ = 4.5), requiring W$_\mathrm{STEM}$ to be substantially wider than that of a segmented electrode fabricated on a Qz handle to maintain RF and optical index matching \cite{yang_loss_2006}. This allows more current to flow in the interaction region, and consequentially reduction in RF loss with segmented electrodes on an Si handle is not seen to the same extent as when using a Qz handle and a very narrow W$_\mathrm{STEM}$ \cite{li_analysis_2004,kharel_breaking_2021}. Regardless, the segmented electrodes reported in this manuscript eliminate the need for a UV15 cladding, which nearly doubles RF loss over 100 GHz \cite{nelan_compact_2022,ahmed_high-efficiency_2020}. Through precise index matching, the EO modulator then shows a 3-dB EO bandwidth, over twice that of a UV15-clad device reported in previous work \cite{nelan_compact_2022}. The aerial view of the EO modulator is shown in Fig. \ref{fig:overview}. (d), with high-resolution images of the probe-transition, interaction region and thermal bias heaters in Fig. \ref{fig:overview}. (e, f, g), respectively. Fabrication begins on a 300 nm X-cut TFLNOI slab procured from NanoLN™. The material stack, from top to bottom is 500 nm of plasma-enhanced chemical vapor deposition (PECVD) SiO$_2$, 100 nm of PECVD SiN$_\mathrm{x}$, which defines the strip-loaded waveguide, 300 nm of X-cut LiNbO$_\mathrm3$, and 4.7 $\mu$m of thermal SiO$_2$ atop a 500 $\mu$m Si handle. All optical structures are defined with electron-beam lithography, while the electrodes are defined with a laser writer. The electrodes are 1 $\mu$m electroplated Au. The thickness and width of the  SiN$_\mathrm{x}$ waveguide is 100 nm and 2 $\mu$m, respectively. With this, roughly 69 $\%$ of the optical mode is confined in the LiNbO$_\mathrm3$ layer. The cross section of the waveguide and electrodes in the interaction region can be seen in Fig. \ref{fig:overview}. (b), a top-down view of the interaction region with dimensions labeled can be seen in Fig. \ref{fig:overview}. (c). The EO modulator employs a 1x2 MMI splitter at the beginning and a 2x2 MMI splitter/combiner at the end to produce two outputs with inversely-proportional optical intensities depending on the applied RF signal. Design, fabrication and characterization of these optical structures is discussed in previous work \cite{nelan_compact_2022,ahmed_subvolt_2020}.

\section{Results and Discussion}

\begin{figure}[ht]
\centering
{\includegraphics[width=\linewidth]{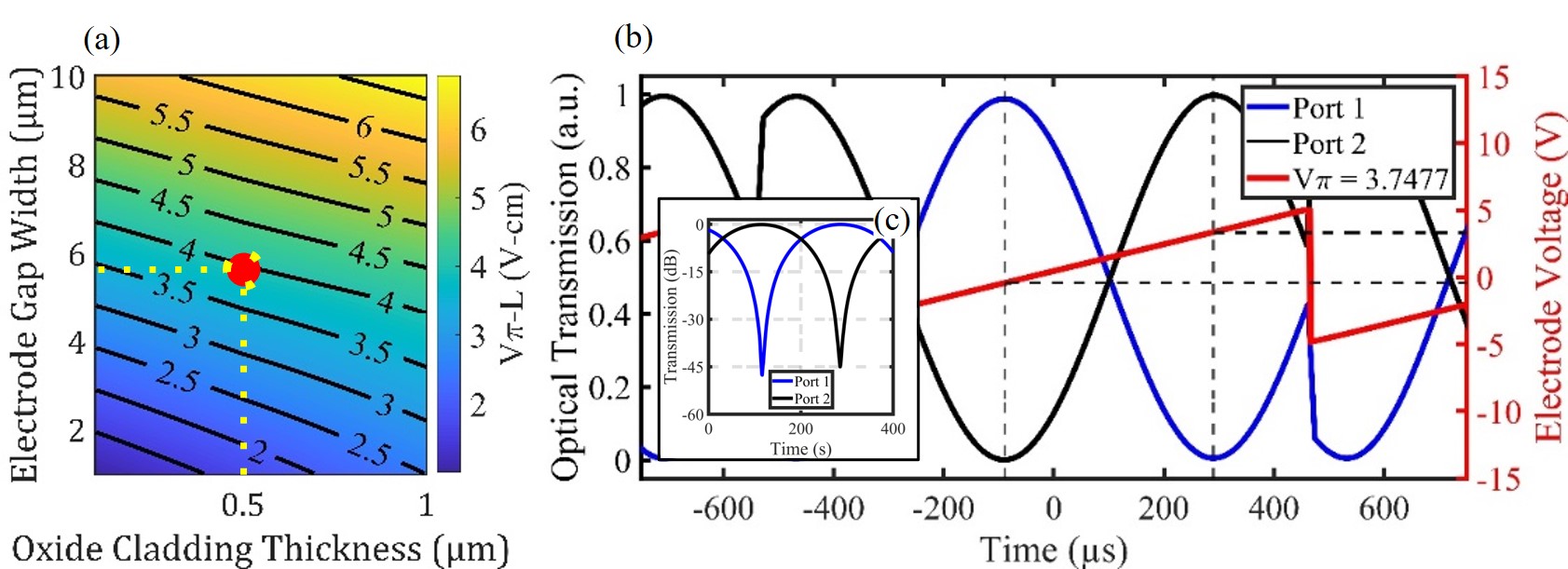}}
\caption{(a) Heat-map comparing the effect of electrode W$_\mathrm{GAP}$ and T$_\mathrm{SIO2}$ on V$\pi$-L. (b) The measured DC-V$\pi$ of the device. The red trace shows applied electrode voltage, the blue and black traces show normalized optical intensity at output ports 1 and 2, respectively. (c) The measured ER of the device at both output ports.}
\label{fig:vpidata}
\end{figure}

\begin{figure*}[h]
\centering
{\includegraphics[width=\linewidth]{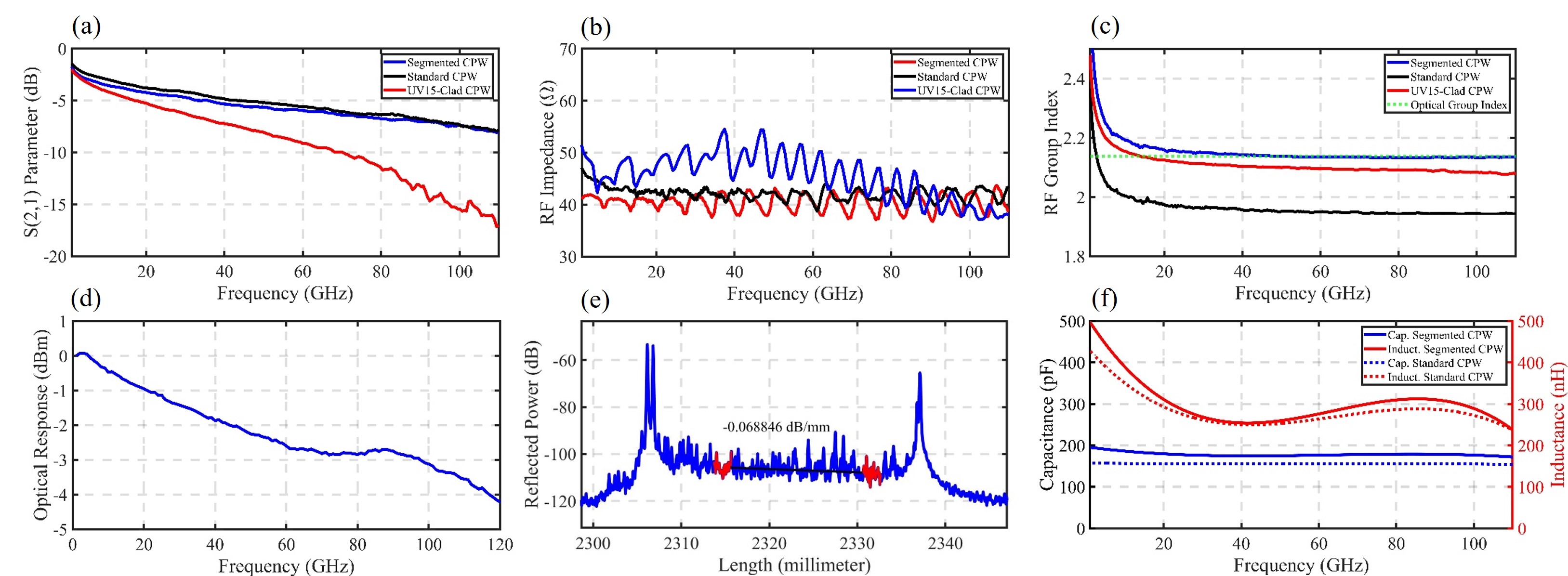}}
\caption{(a) Measured S(2,1) response of the CPW designs. (b) Measured impedance of the CPW designs. (c) Measured RF group index of the CPW designs. (d) Measured electro-optic response from 1-120 GHz. (e) Measured propagation loss and back-scatter of the optical waveguide. (f) Capacitance and inductance of the CPW designs.}
\label{fig:hfdata}
\end{figure*}

A Keysight PNA-X with range extenders and W-band ground-signal-ground probes is used to measure RF scattering parameters of the different electrode designs. In Fig. \ref{fig:hfdata}. (a), it is shown that the UV15-cladding nearly doubles the RF loss of the electrode at 120 GHz where absorption loss in the UV15 becomes dominant. There is a rapid drop-off in S(2,1) from DC to roughly 1 GHz from a low-frequency impedance mismatch. On the Si handle, with its high-microwave absorption tangent (tan$\delta$ = 10$^{-2}$), dielectric loss remains dominant over conductive or radiative losses \cite{Chudpooti_2021}. Furthermore, considering $\epsilon_\mathrm{Si}$ = 11.7, W$_\mathrm{STEM}$ must be relatively wide to properly match $n_{RF}$ and $n_{og}$ which does not significantly reduce current bunching in the interaction region when compared to a standard CPW design \cite{kharel_breaking_2021}. As such, it is shown that there are no outstanding improvements to RF loss when compared outright to a standard CPW. Nevertheless, the RF loss is halved compared to an index-matched, UV15-clad CPW.

Z$_{l}$ and $n_{RF}$ are extracted from the measured scattering parameters and presented in Fig. \ref{fig:hfdata}. (b) and (c), respectively. While $n_{RF}$ and $n_{og}$ are extremely well matched at high frequencies, there is a +/- 10 $\Omega$ variation in Z$_{l}$ along the frequency sweep. Considering the effect of Z$_{l}$ on the AC-V$\pi$ and EO S(2,1) penalty, following $(Z_{l}+50\Omega) / 2Z_{l}$ and 20log($2Z_{l}/(Z_{l}+50\Omega)) $, respectively, a 10 $\Omega$ mismatch results in a 0.47 V increase in V$\pi$ and a 1 dB reduction in EO S(2,1) \cite{ghione_2009,kharel_breaking_2021}. The DC EO response and ER of the device are presented in Fig. \ref{fig:vpidata}. (b) and (c), respectively. The DC V$\pi$ measures roughly 3.75 V, while the ER is >45 dB at each port. Both output ports show an output intensity inversely proportional to the other. The measured sideband power of the modulator from 1 GHz to 120 GHz is shown in Fig. \ref{fig:hfdata}. (d). Here, the device is operating at quadrature bias, and the sideband power is normalized to the carrier and RF input power. The measured 3-dB response of the modulator is roughly 95 GHz using 1 GHz as the reference frequency. A flattening of the sideband response can be seen above 70 GHz which corresponds to the gradual improvement in both $Z_{l}$ and $n_{RF}$. This effect may be related to the total length of each segment, but has not been investigated \cite{li_analysis_2004}. The EO modulator shows a fiber-to-fiber insertion loss of roughly 7.47 dB using OZ Optics lensed fibers with a mode field diameter (MFD) of 2.5 $\mu$m. A reflectometer/light-wave analyzer is then used to determine the back-scatter and propagation loss in the device, shown in Fig. \ref{fig:hfdata}. (e). Optical propagation loss is roughly 0.69 dB/cm, which is consistent with previous work. The total length of the reference optical waveguide is roughly 30 mm which gives a total propagation loss of 2.07 dB, while an additional 2.7 dB of optical power is lost at each facet.

\section{Conclusion}

In this work, a high-speed, dual-output LiNbO$_\mathrm3$-SiN$_\mathrm{x}$ electro-optic MZM using segmented electrodes is fabricated on a Si handle without the need for substrate removal to demonstrate the feasibility of a micro-structured CPW in this material platform. Micro-structured electrodes provide tunability of $n_{RF}$, and drastically reduce the RF loss compared to a UV15-clad electrode, increasing EO bandwidth. The use of a Si handle allows the use of larger wafer sizes to increase yield in production environments, lends itself to processing in established Si foundries, and reduces the cost of upgrading existing systems with this new technology so that it may one day come to fruition in global telecommunications systems. Without the need for substrate removal or an index-matching epoxy, the device is extremely durable and can reliably operate in a myriad of environmental conditions. The modulator has an interaction region of 10 mm, shows a DC half wave voltage (V$\pi$) of 3.75 V, an ultra-high ER of over 45 dB at both output ports and a fiber-to-fiber insertion loss of 7.47 dB. The measured 3 dB bandwidth is roughly 95 GHz. Finally, there is a path forward for further improvement of RF performance by optimization of the "T"-structure to reduce current crowing in the interaction region.

\section{Backmatter}

\begin{backmatter}

\bmsection{Acknowledgments} This manuscript (AFRL-2022-2292) was supported in part under AFRL contract FA8650-19-C-1027. The authors gratefully acknowledge the support of UTA16-001296. The views and conclusions contained herein are those of the authors and should not be interpreted as necessarily representing the official policies or endorsements, either expressed or implied, of Air Force Research Laboratory, the Department of Defense, or the U.S. Government.

\bmsection{Disclosures} The authors declare no conflicts of interest.

\bmsection{Data Availability Statement} Data underlying the results presented in this paper are not publicly available at this time but may be obtained from the authors upon reasonable request.

\end{backmatter}

\section{References}

\bibliography{SegmentedLetterSources.bib}

\bibliographyfullrefs{SegmentedLetterSources.bib}

\end{document}